\documentclass[11pt]{article}

\usepackage{amsmath}
\usepackage{graphicx}
\usepackage[squaren]{SIunits}
\usepackage[unicode]{hyperref}
\usepackage{cite}

\newcommand{\gfrac}[2]{\displaystyle\frac{#1}{#2}}
\newcommand{\dd}{\mbox{d}}

\def\threshold{\text{threshold}}

\begin{document}

\title{A 5D, polarised, Bethe-Heitler event generator for $\gamma \to \mu^+\mu^-$ conversion}

\author{D.~Bernard, \\
LLR, Ecole Polytechnique, CNRS/IN2P3, 91128 Palaiseau, France}

\maketitle 

\begin{abstract}
I describe a five-dimensional, polarised, Bethe-Heitler event
generator of $\gamma$-ray conversions to $\mu^+\mu^-$, based on a
generator for conversion to $e^+e^-$ developed in the past.
Verifications are performed from close-to-threshold to high energies.
\end{abstract}

{\em keywords}:

gamma rays, 
pair conversion, 
muon pairs, 
gamma factory, 
event generator,
Bethe-Heitler,
Geant4

\section{Introduction}

In the past I wrote an event generator of $\gamma$-ray
conversions to an $e^+e^-$ pair \cite{Bernard:2013jea} that we used
for the simulation of the HARPO experiment
\cite{Gros:2017wyj,Bernard:2018jql}.
The generator samples the exact, that is, five-dimensional,
Bethe-Heitler differential cross section.
Later I developed a new version of that code, the fortran model
\cite{Bernard:2018hwf} of an algorithm appropriate to event generation
in Geant4.
The Geant4 toolkit is used for the detailed simulation of 
scientific experiments that involve the interaction of ``elementary'' 
particles with matter, in particular with a detector 
\cite{Agostinelli:2002hh,Allison:2016lfl,Apostolakis:2015elm}.
The generator has been made available as the G4BetheHeitler5DModel
physics model of Geant4 since release 10.5
\cite{Semeniouk:2019cwl,VIvanchenko:2019},
see also SubSubSect. 6.5.4 of \cite{10.5}.

A Gamma Factory project is being proposed at CERN in the photon energy range
$1 < E < 400\,\mega\electronvolt$
\cite{Krasny:2018alc}, with photon intensities orders of magnitude
above that presently available.
Such a facility could be a powerful source of muons by gamma-ray
conversions to $\mu^+\mu^-$, and therefore also a powerful source of neutrinos.
A precise simulation of photon conversion to muons pairs was therefore
intensely desirable \cite{VIvanchenko:2019:2}.

The $\gamma \to \mu^+\mu^-$ physics model presently available in
Geant4, {\em G4GammaConversionToMuons} is based on the algorithm
described in \cite{Burkhardt:2002vg} and in Sect. 6.7 of \cite{10.5} and
in references therein.
The calculation makes use of high-energy approximations and the model
has been verified above incident photon energies of 10\,GeV.

In this note I characterise a modified version of the fortran model of
conversions to $e^+e^-$ documented in \cite{Bernard:2018hwf} for
conversions to $\mu^+\mu^-$ and that is valid down to threshold, either for
\begin{itemize}
\item conversions 
\begin{itemize}
\item on a nucleus (nuclear conversion)
 $\gamma ~ Z ~ \to ~ \mu^{+}\mu^{-} ~ Z $;

\item on an electron (triplet conversion)
 $\gamma ~ e^{-} ~ \to ~ \mu^{+}\mu^{-} ~ e^{-} $.
\end{itemize}

\item on a target in an atom, or on an isolated target (``QED'')

\item for linearly polarised or non-polarised incident photons.
\end{itemize}

The differential cross section for conversions of non-polarised
photons was obtained by Bethe \& Heitler \cite{Bethe-Heitler}, and
that to linearly polarised photons by
\cite{BerlinMadansky1950,May1951,jau}\footnote{The polarised
 differential cross section first obtained by
 \cite{BerlinMadansky1950} was put in Bethe-Heitler form by
 \cite{May1951}, after which a misprint was corrected by
 \cite{jau}.}. 
The final state is five-dimensional, even when the recoiling target
cannot be observed.
It, the final state, can be defined by the polar angles $\theta_+$ and
$\theta_-$, and the azimuthal angles $\phi_+$ and $\phi_-$, of the
negative lepton ($-$) and of the positive lepton (+), respectively, and
the fraction $x_+$ of the energy of the incident photon carried away
by the the positive lepton, $x_+ \equiv E_+/E$
(Table\,1 of \cite{Bernard:2018hwf}).
$q$ is the momentum ``transferred'' to the target.
In case conversion takes place in the field of an
isolated nucleus or electron, the Bethe-Heitler expression is used,
while when the nucleus or the electron is part of an atom, the
screening of the target field by the other electrons of the atom is
described by a form factor, a function of $q^2$ \cite{Mott:1934}
(nuclear) or \cite{WheelerLamb1939} (triplet).

Please note that in contrast with Bethe-Heitler \cite{Bethe-Heitler},
I do not assume that the energy carried away by the recoiling target
be negligible (compare the differential element of
eq. (20) of \cite{Bethe-Heitler} to that of
eqs. (1)-(3) of \cite{Bernard:2018hwf}).

The differential cross section diverges at small $q$ and, for
high-energy leptons, in the forward direction ($\theta \approx 0$).
The $1/q^4$ divergence  is a particularly
daunting nuisance as the expression for $q$ involves several of the
variables that describe the final state.
The inconvenience for event generation is overcome as is usual in 
particle physics, by performing each step of the conversion
(gamma-target $\to$ recoiling target-pair; pair $\to$ two leptons) in
its own centre-of-mass frame (Sect. 3 of \cite{Bernard:2018hwf} and
references therein).

The problem is then reduced to using the Von Neumann
acceptance-rejection method (Sect. 40.3 of \cite{Tanabashi:2018oca})
from a mock-up probability density function (pdf) that is the simple
product of five independent pdfs, of five judiciously chosen variables
$x_i, i=1\cdots 5$.

\section{Bounds}
\label{sec:bounds}

\begin{figure}[tbh]
 \begin{center}
 \includegraphics[width=0.8\linewidth]{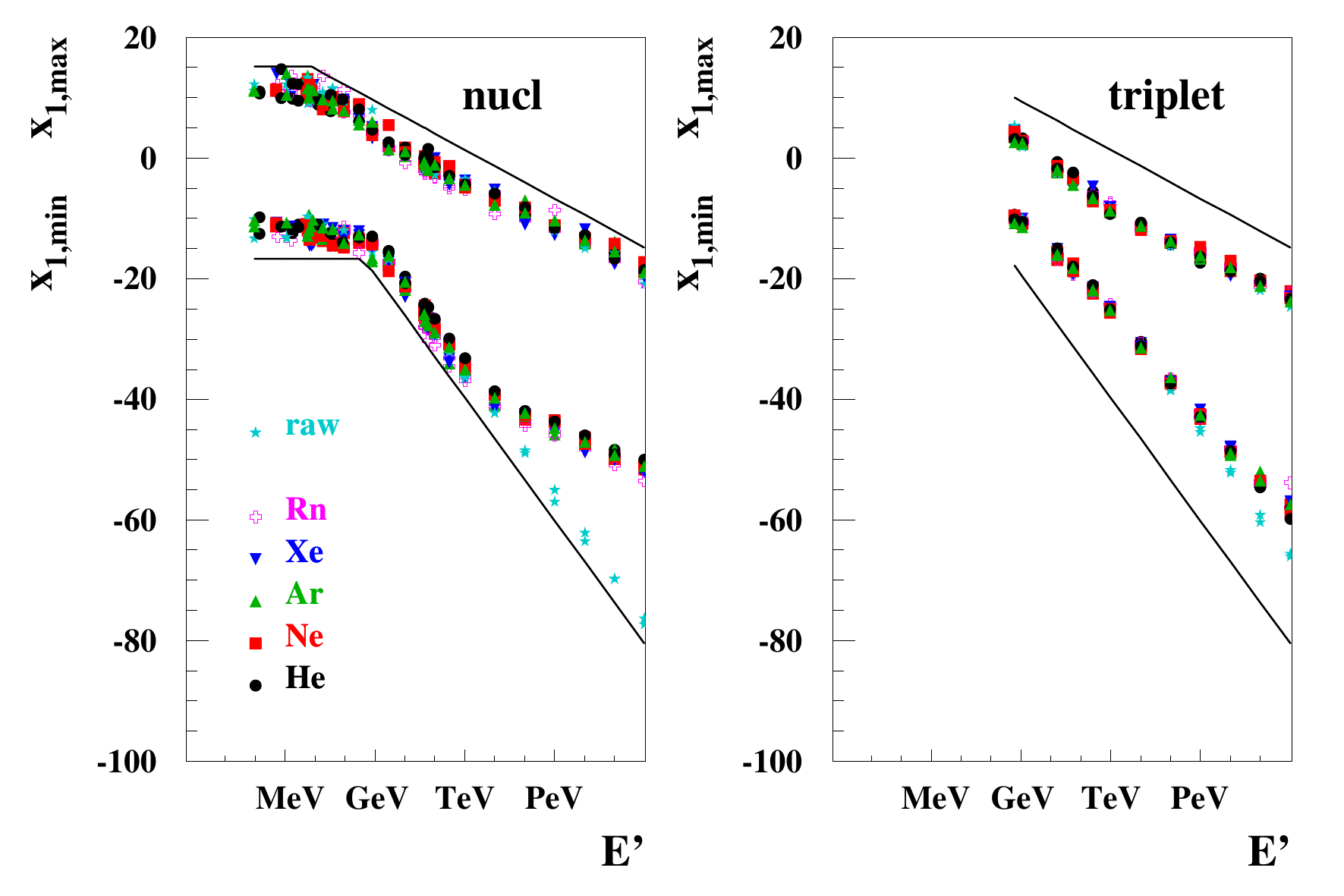}
\caption{Variation with the available energy $E'$ of the maximum and
 of the minimum values for variable $x_1$, for photon conversions to
 $\mu^+\mu^-$.
 Left: nuclear conversion; Right: triplet conversion.
``raw'' isolated charged target (star), and the following atoms:
helium (bullet),
neon (square),
argon (upper triangle),
xenon (down triangle) and
radon (plusses).
The lines denote the bounds that are used by the generator.
$P=1$ and $P=0$ samples are plotted for each photon-energy value.
\label{fig:bounds}
}
 \end{center}
\end{figure}

The bounds of $x_i, i=2\cdots 5$ are pretty straightforward and do not
depend on the mass of the leptons, see Table 3 of
\cite{Bernard:2018hwf}.
For $x_1$, though, the bounds must be provided and they depend on the
energy of the incident photon, see Fig. 1 of \cite{Bernard:2018hwf}
for conversions to $e^+e^-$.
The $x_1$ bounds for conversions to $\mu^+\mu^-$ are similar and are
shown in Fig. \ref{fig:bounds}.

\begin{figure*}[bth] 
\includegraphics[width=0.49\linewidth]{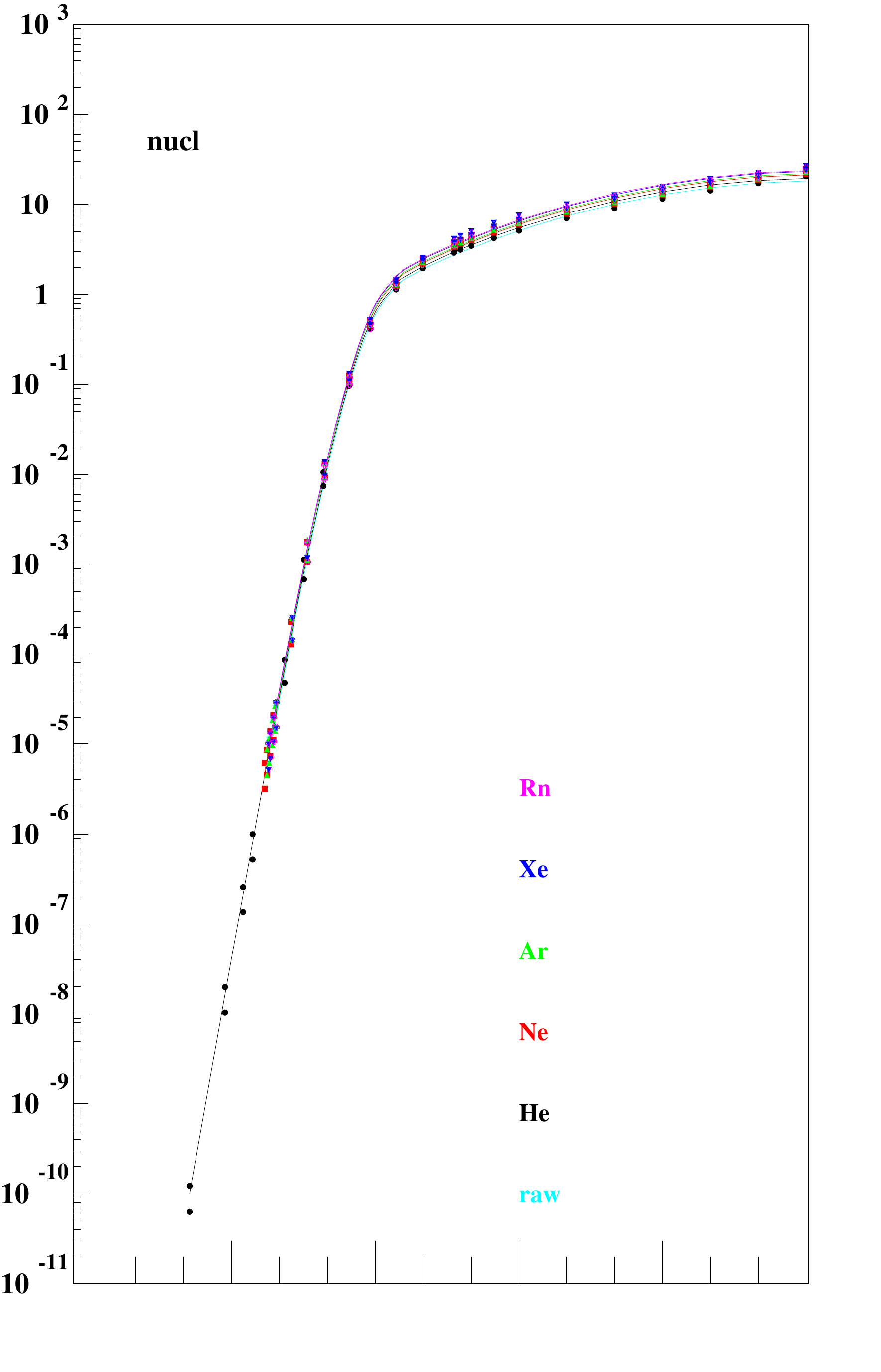}
\includegraphics[width=0.49\linewidth]{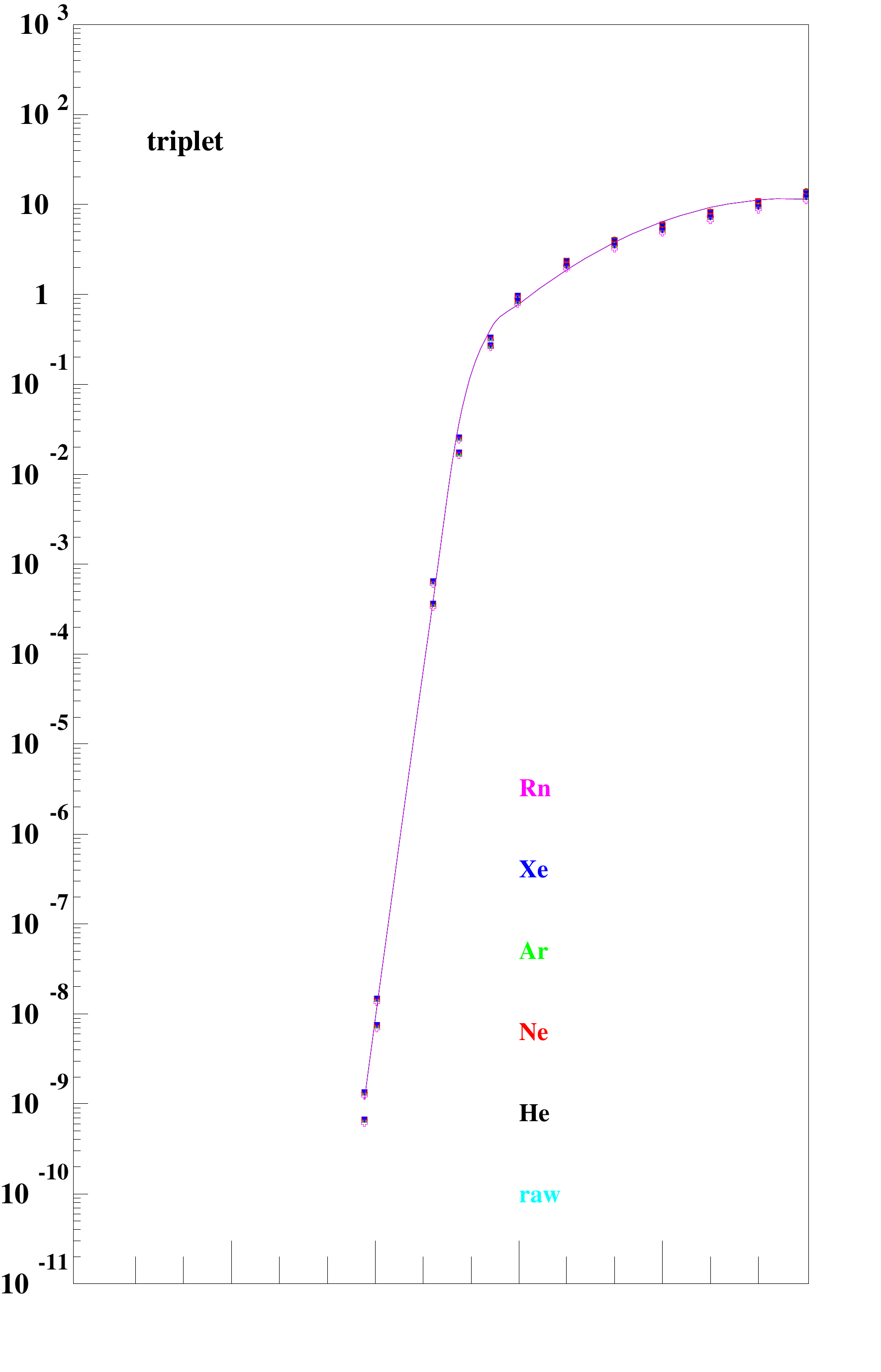}

\vspace{-1.1cm}
\includegraphics[width=0.49\linewidth]{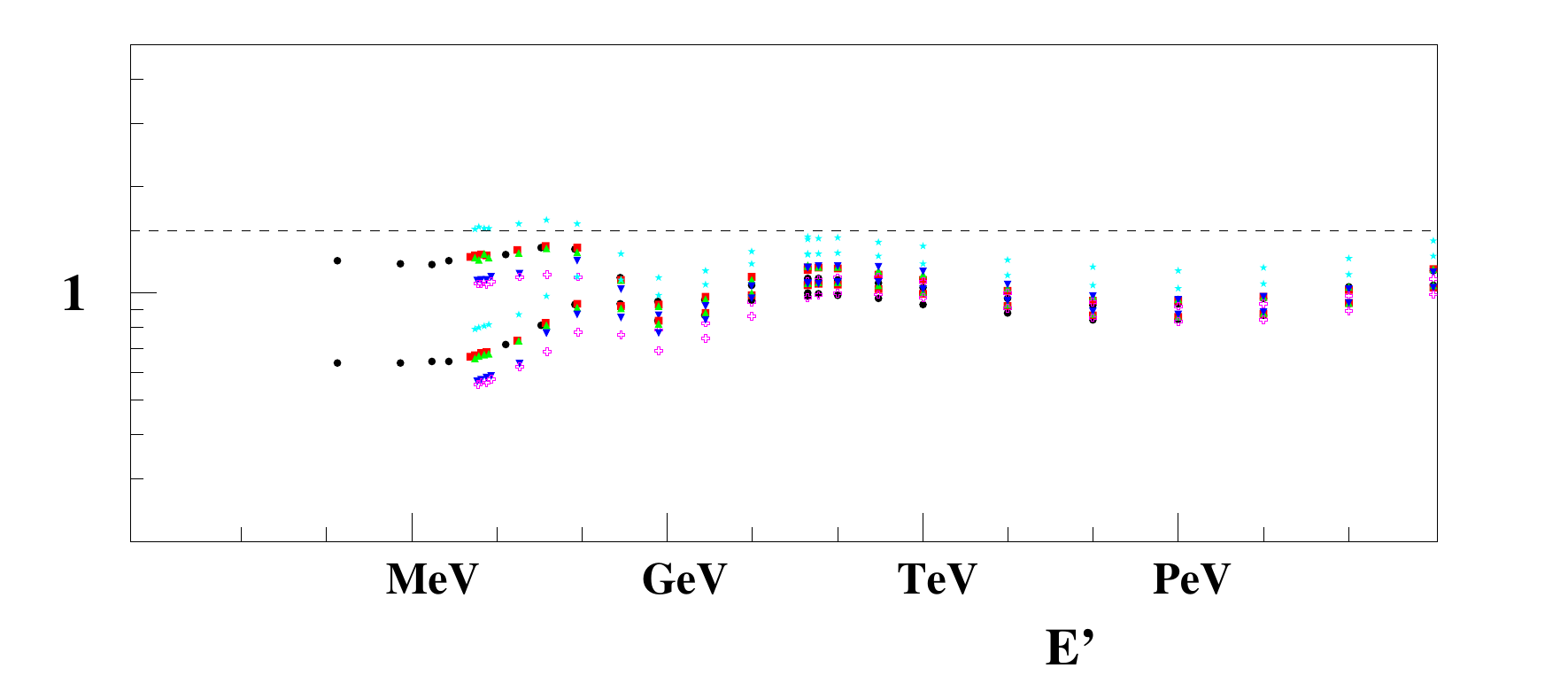}
\includegraphics[width=0.49\linewidth]{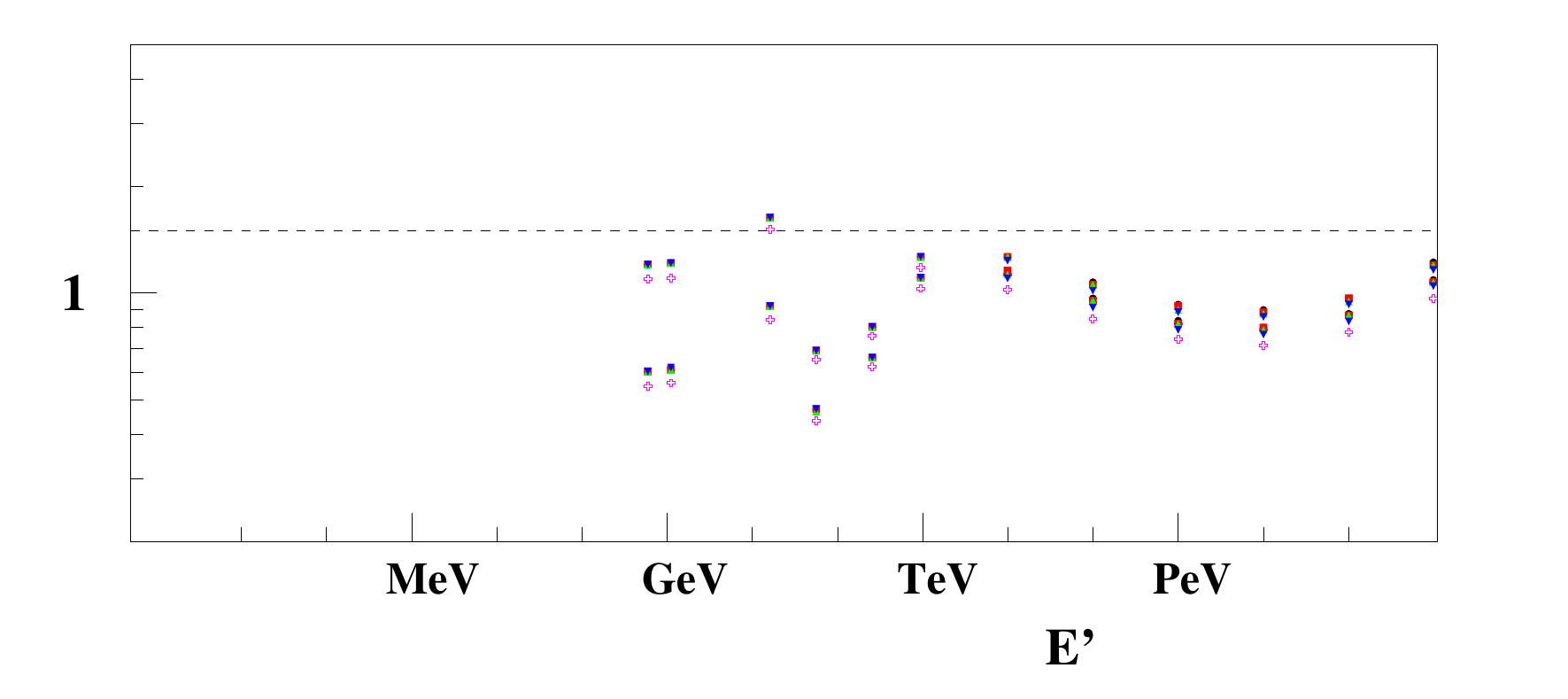}
\caption{Top: Variation of the maximum value of the pdf
 (arbitrary units) with the available energy $E'$.
Left: nuclear conversion; Right: triplet conversion.
``raw'' isolated charged target (star), and the following atoms:
helium (bullet),
 neon (square),
argon (upper triangle),
 xenon (down triangle) and
radon (plusses).
The curves denote the fit to the present data of the
expression of eq. (9) of \cite{Bernard:2018hwf}).
Bottom: residues of the fit, with the 1.5 safety factor marked with a
dashed line.
$P=1$ and $P=0$ samples are plotted for each photon-energy value.
\label{fig:max:diffcross}
 }
\end{figure*}

For better readability, energy-variation plots are presented as a
function of the available energy 
\begin{equation}
 E' = E - E_{\threshold},
\end{equation}
where the energy threshold is

\begin{equation}
E_{\threshold} = 2 (m_\mu^2 / M + m_\mu)
 ,
\end{equation}
where $M$ is the target mass.
\begin{itemize}
\item 
For triplet conversion to $\mu^+\mu^-$, $E_{\threshold}$ is close to 43.9\,GeV.
\item 
For nuclear conversion to $\mu^+\mu^-$, $E_{\threshold}$ ranges from
223.2\,MeV for hydrogen targets down to almost
$2 m_\mu \approx 211.3\,\mega\electronvolt$ for heavy nuclei.
\end{itemize}

The $x_1$ bounds are clearly not quite optimal for triplet conversion
and might be tightened a bit further separately.
The CPU gain for routine Geant4 use would be small though, due to the
$1/Z$ suppression of triplet with respect to nuclear and due to the much
higher threshold energy.

\section{Maximum value of the differential cross section}
\label{sec:maxdif}

The acceptance-rejection method needs the knowledge of the maximum
value of the differential cross section, so as to determine a
value of the constant $C$ (see Sect. 5 of \cite{Bernard:2018hwf} or
Sect. 40.3 of \cite{Tanabashi:2018oca}).
The variation of the maximum is shown in Fig. \ref{fig:max:diffcross}.
As the maxima for all target masses were obtained from the same set of
photon energies, and as the threshold varies with target mass for
nuclear conversion, the available energy $E'$ also varies with target
mass.
In the nuclear conversion plot, for example, for
$E = 217.40\,\mega\electronvolt$ photon conversion on helium, 
$E_{\threshold}$ is $ \approx 217.26\,\mega\electronvolt$, that is, 
$E' \approx 140\,\kilo\electronvolt$.
The parametrisation of the maximum is obtained with the same expression
as for $e^+e^-$ (eq. (9) of \cite{Bernard:2018hwf}) with a different set
of parameters.

In practice, for safety, $C$ is enlarged by a factor of 1.5 with
respect to what can be seen on Fig. \ref{fig:max:diffcross}.

\section{Distributions}
\label{sec:dist}

Figs. \ref{fig:bounds} -- \ref{fig:asym} are based on samples of
$10^5$ simulated events, and
Figs. \ref{fig:xplus} -- \ref{fig:ouverture} on samples of $10^6$
simulated events.
``Raw'' nuclear samples, that is, simulations of conversions of an
isolated target (``QED'') have been produced assuming an argon nucleus
mass.

\subsection{Polarisation asymmetry}

\begin{figure}[ht] 
 \includegraphics[width=0.49\linewidth]{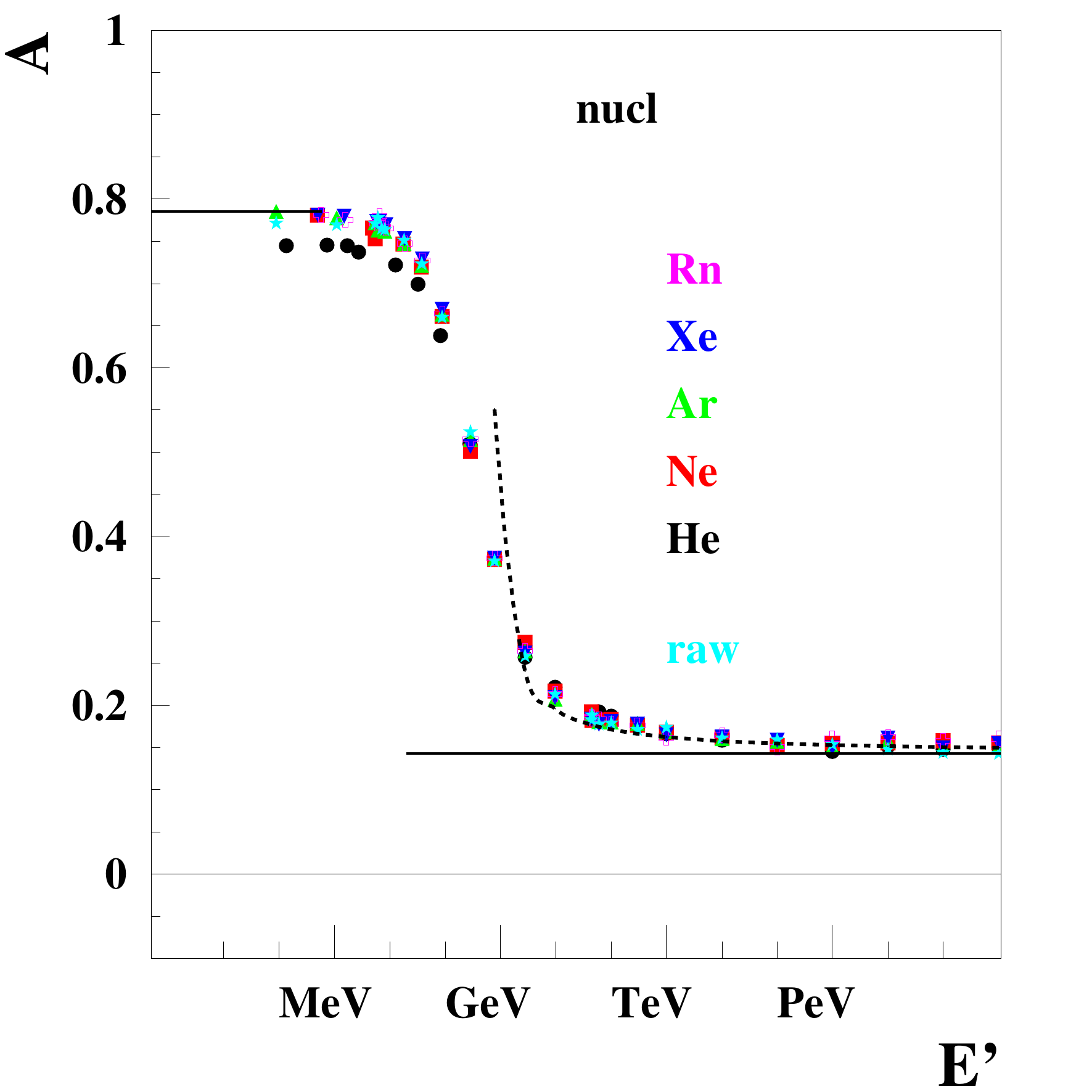}
 \includegraphics[width=0.49\linewidth]{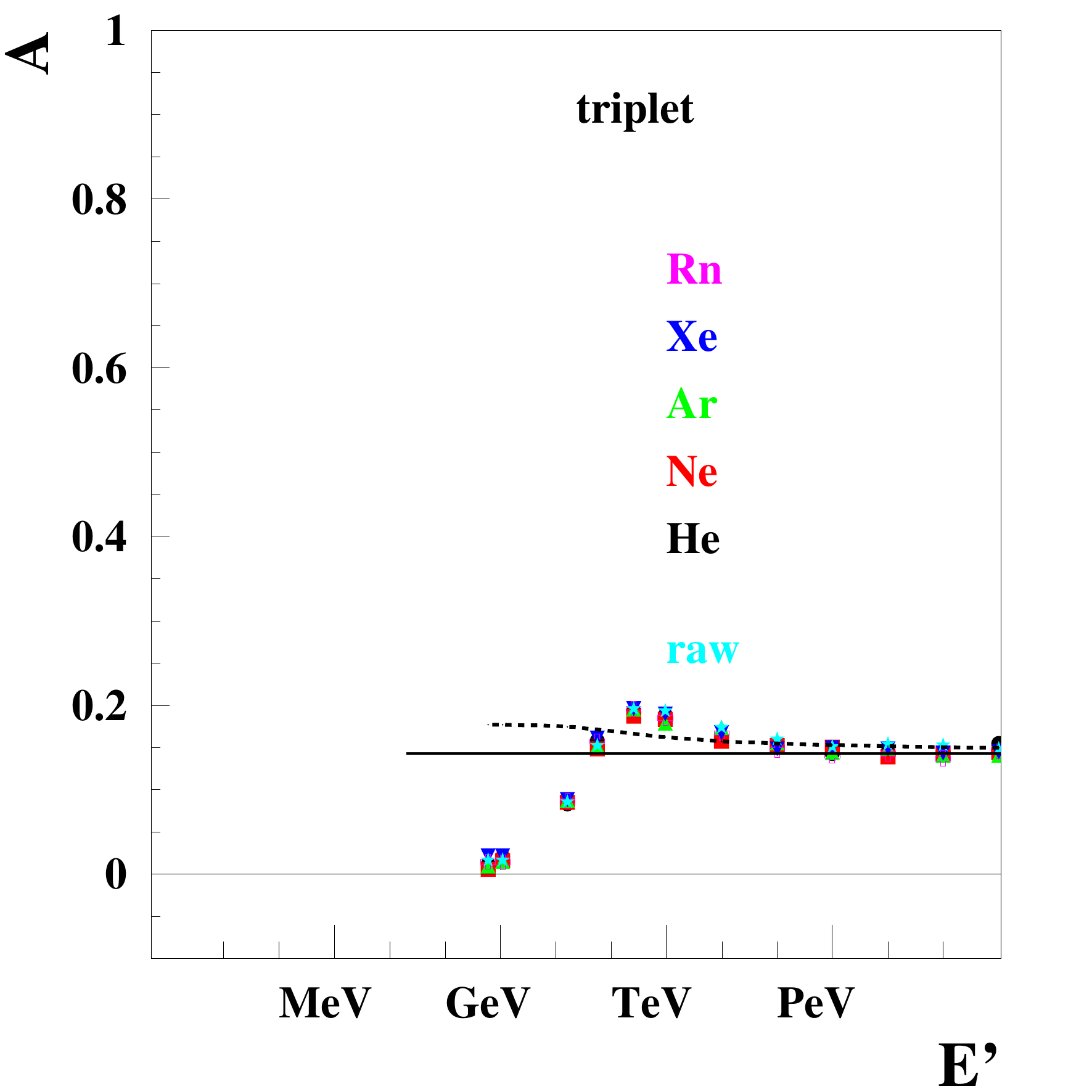}
 \put(-480,190){\boldmath $\gfrac{\pi}{4}$}
 \put(-255,68){\boldmath $\gfrac{1}{7}$}
 \put(-15,68){\boldmath $\gfrac{1}{7}$}
\caption{Polarisation asymmetry calculated on simulated event samples
 as a function of $E'$.
Left: nuclear conversion.
Right: triplet conversion.
``raw'' isolated charged target (star), and the following atoms:
helium (bullet),
 neon (square),
argon (upper triangle),
 xenon (down triangle) and
radon (plusses).
The horizontal lines denote the low- and high-energy approximations
of $\pi/4$ and $1/7$, respectively.
The dashed curves denote the Boldyshev-Peresunko high-energy
approximation\,\cite{Boldyshev:1972va} and shown here in eq. (\ref{eq:sig:HE}).
\label{fig:asym}
 }
\end{figure}

The measurement of the polarisation angle $\varphi_0$ and of the
linear polarisation fraction $P$ of a gamma-ray beam can be performed
by the analysis of the distribution of the event azimuthal angle
$\varphi$
\begin{equation}
\gfrac{\dd N}{\dd \varphi} \propto 
\left(
1 + A \, P \cos[2(\varphi - \varphi_0)]
\right),
\label{eq:modulation}
\end{equation}

Figure \ref{fig:asym} shows the polarisation asymmetry, $A$, obtained
from the ($P=1$) samples, defining the event azimuthal angle as the
bisector of the electron and of the positron azimuthal angles,
$\varphi \equiv (\phi_++\phi_-)/2$ \cite{Gros:2016dmp}, and using the
moments' method \cite{Bernard:2013jea,Gros:2016zst,Gros:2016dmp}.
 
Results agree nicely with the Boldyshev-Peresunko asymptotic
expression\,\cite{Boldyshev:1972va} at high energy:
\begin{eqnarray}
A \approx \gfrac
{\left(\gfrac{4}{9}\right)\log{\left(\gfrac{2 E}{m_\mu c^2}\right)} - \left(\gfrac{20}{28}\right)}
{\left(\gfrac{28}{9}\right)\log{\left(\gfrac{2 E}{m_\mu c^2}\right)} - \left(\gfrac{218}{27}\right)}.
\label{eq:sig:HE}
\end{eqnarray}
At low energy the obtained values agree with the asymptotic value of
$\pi/4$ obtained in \cite{Gros:2016dmp} for nuclear conversion and for
nuclei that are much heavier than a muon, mildly for helium, but tends
to zero for triplet conversion.
 
\subsection{Energy share}

\begin{figure}[tb]
 \includegraphics[width=0.47\linewidth]{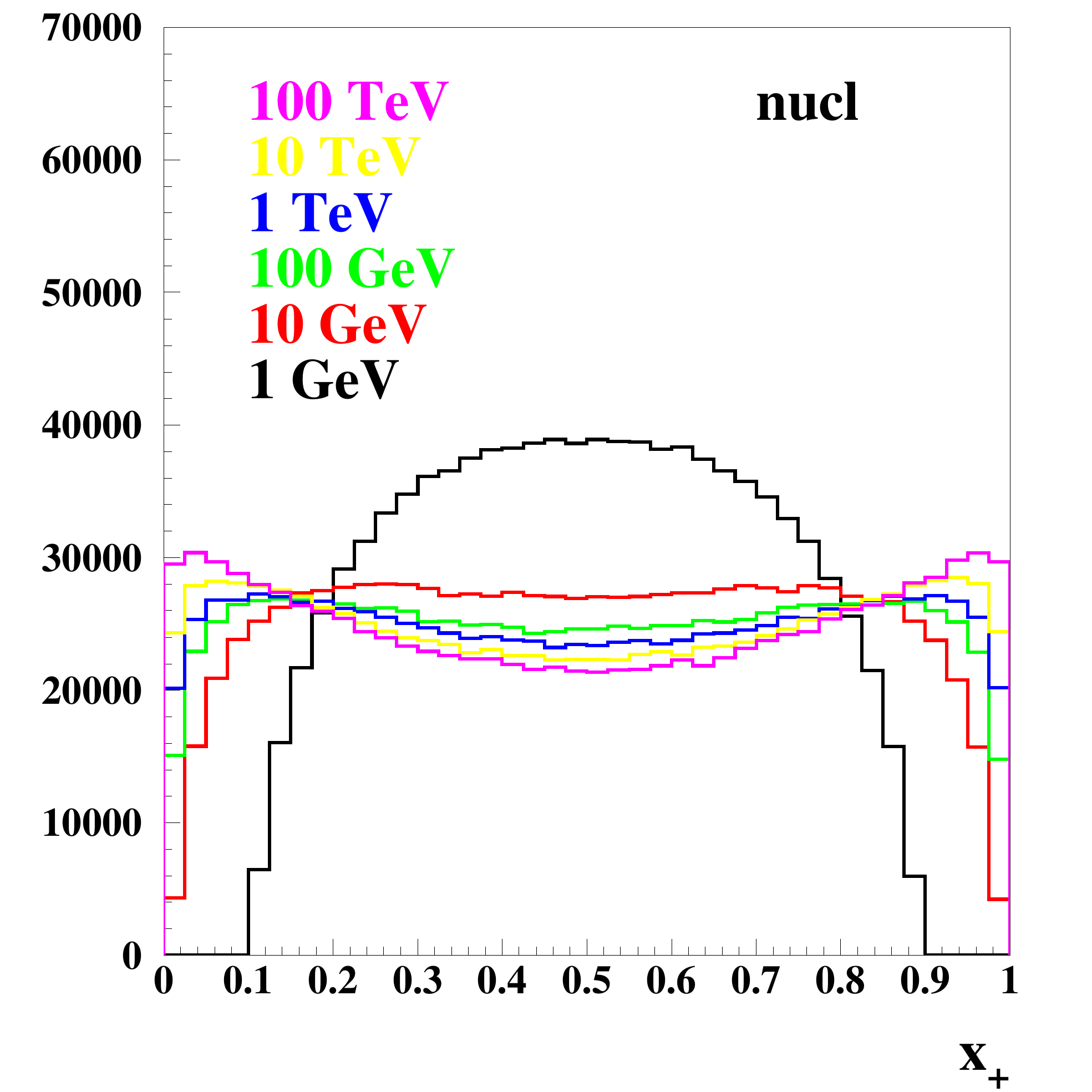}
 \includegraphics[width=0.47\linewidth]{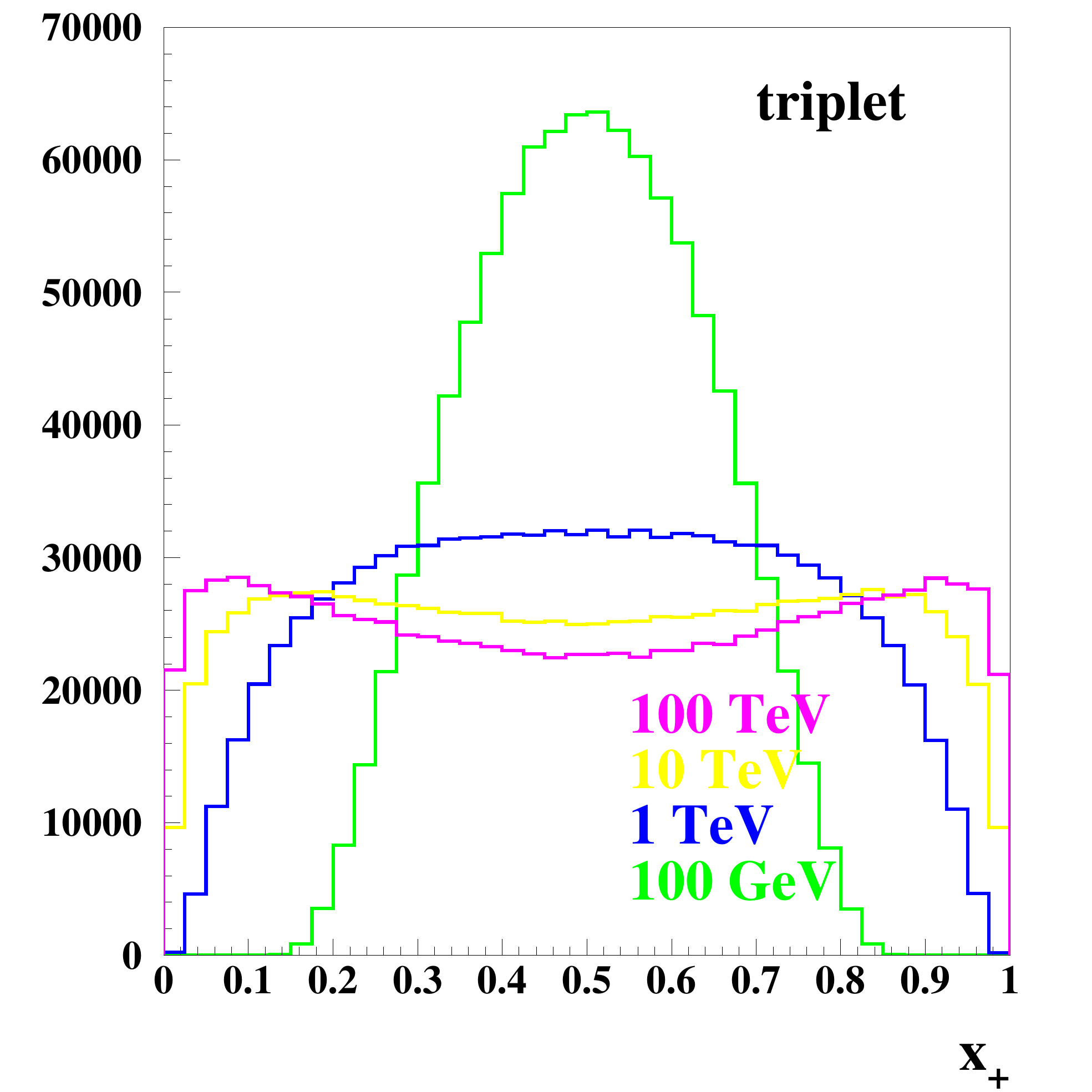}
\caption{Distributions of the fraction of the energy of the incident
photon that is taken away by the positive lepton,
$x_+ \equiv E_+ / E$, for various values of $E$
(conversions on argon). \label{fig:xplus}  }
\end{figure}

The distributions of the fraction $x_+ \equiv E_+/E$ of the energy of
the incident photon that is taken away by the positive lepton is shown
in Fig. \ref{fig:xplus}.
They compare nicely with the published representations of the
singly-differential cross section obtained by integration on all other
variables (Fig. 5 of \cite{Bethe-Heitler}).
 
\subsection{Recoil momentum}

\begin{figure}[ht] 
 \includegraphics[width=0.47\linewidth]{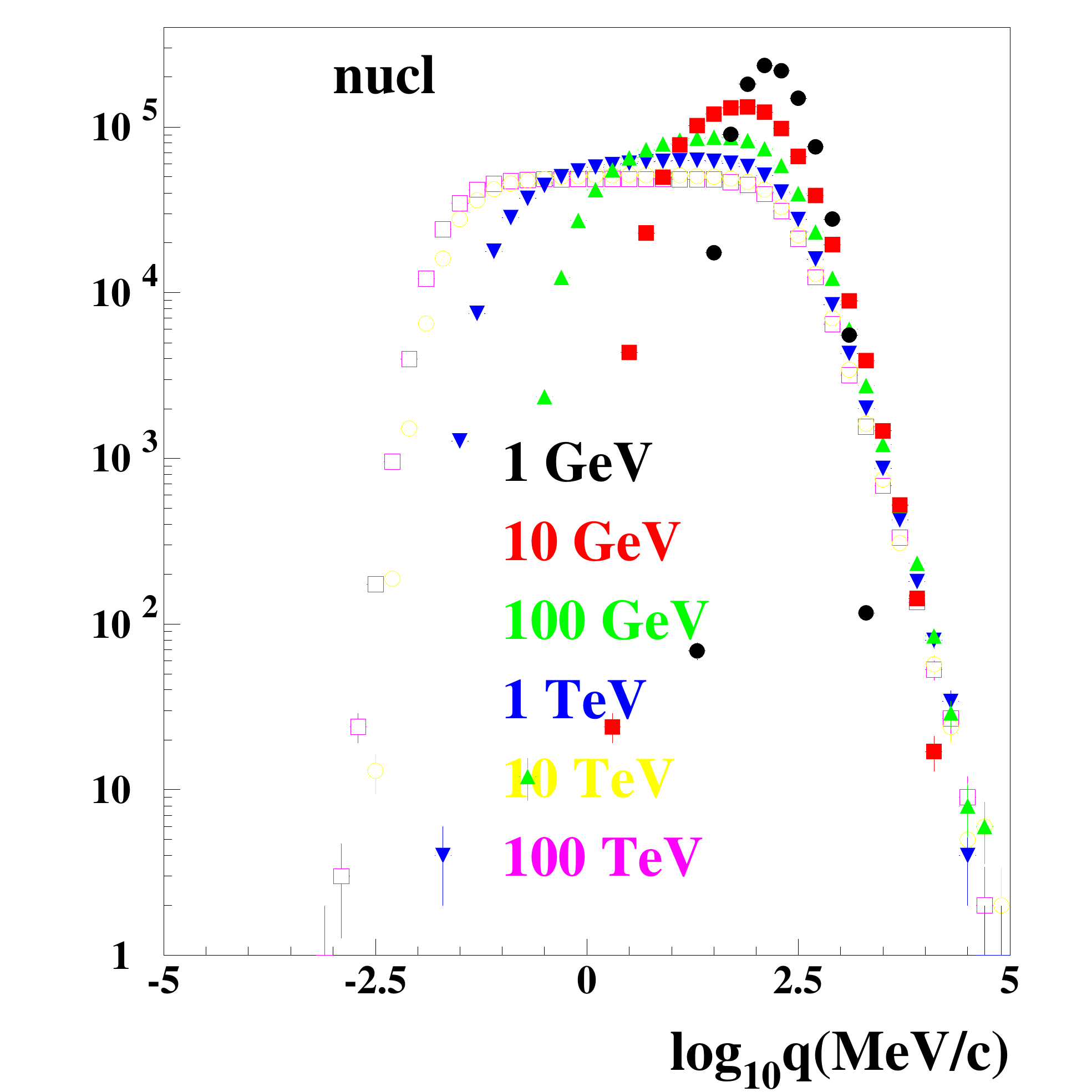}
 \includegraphics[width=0.47\linewidth]{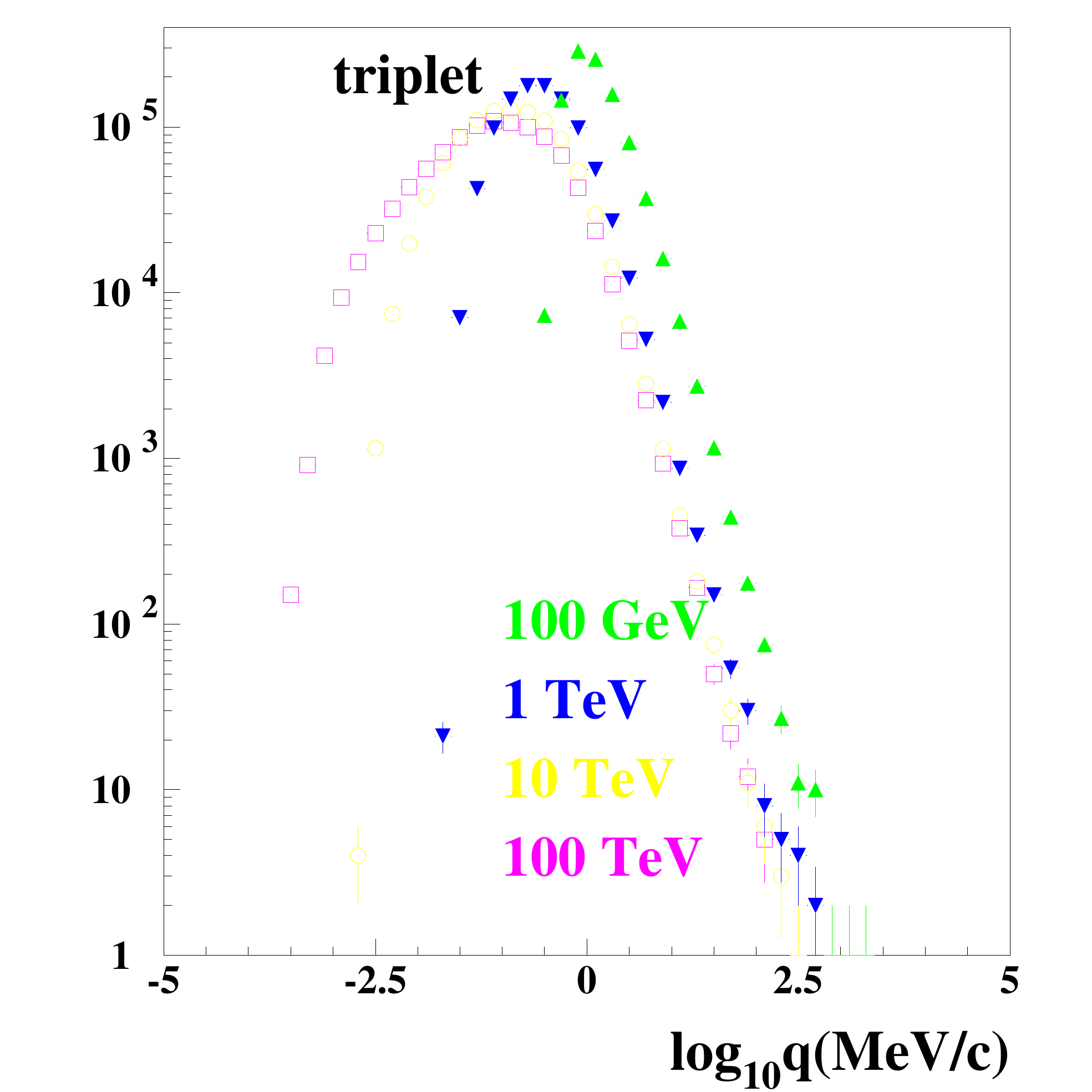}
\caption{($\log_{10}$ of the) recoil momentum spectra for gamma
 conversions on argon atoms.
\label{fig:recoil}
 }

\end{figure}

The distribution of the ($\log_{10}$ of the) recoil momentum is shown in
Fig. \ref{fig:recoil}.

\subsection{Opening angle}

\begin{figure}[bt] 
 \includegraphics[width=0.49\linewidth]{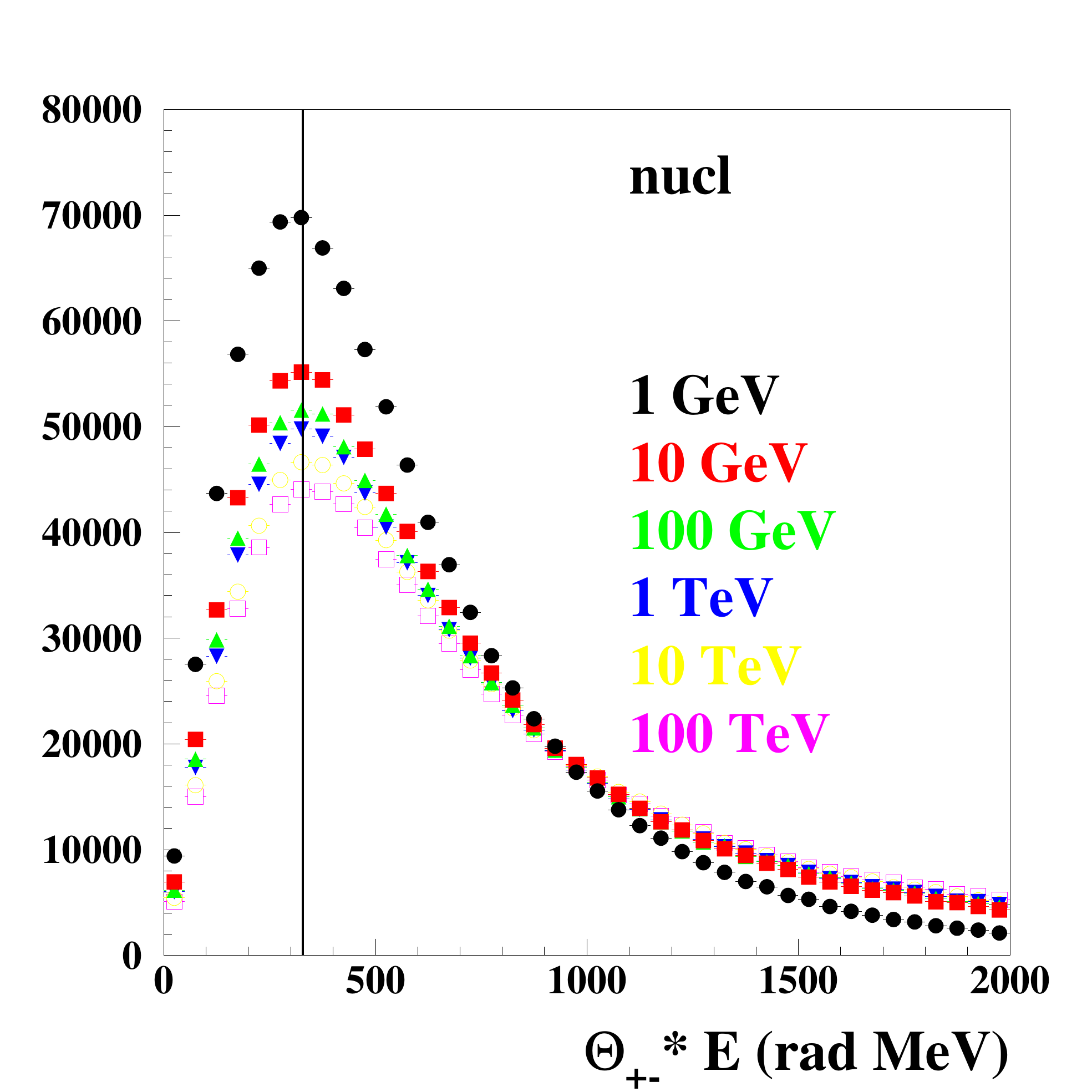}
 \includegraphics[width=0.49\linewidth]{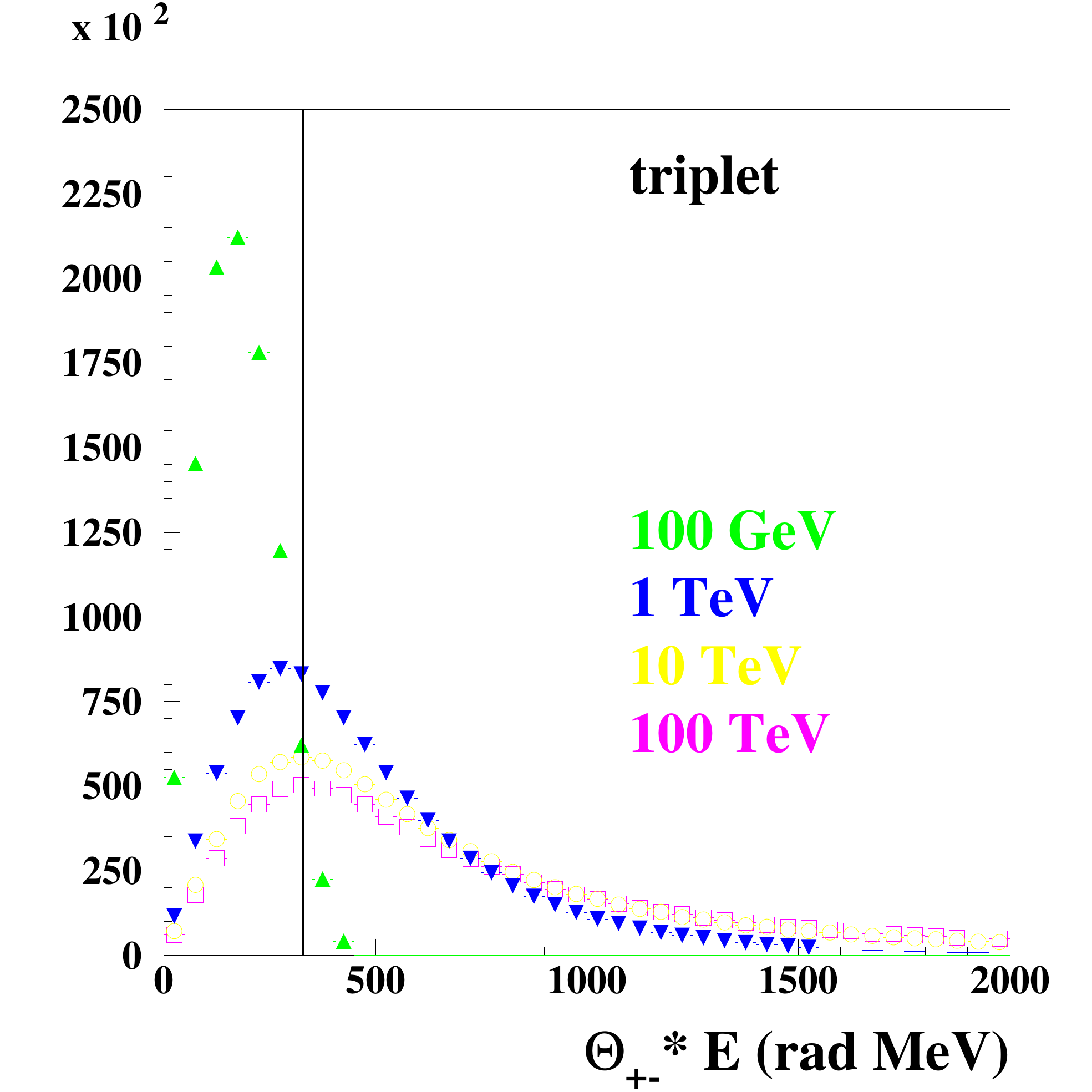}
 \caption{Distributions of the product of the pair opening angle and
 of the photon energy,
 $\theta_{+-} \times E$, for nuclear conversions on argon for
 1\,GeV (bullet),
 10\,GeV (full square),
 100\,GeV (upper triangle),
 1\,TeV (down triangle),
 10\,TeV (circles) and
 100\,TeV (empty squares).
The vertical value shows the most probable value of
$\approx 338.\,\radian\cdot\mega\electronvolt$ computed by Olsen in the
high-energy approximation \cite{Olsen:1963zz}
(see also eqs. (\ref{eq:Olsen})-(\ref{eq:Olsen:mu})).
\label{fig:ouverture}
 }
\end{figure}

Figure \ref{fig:ouverture} shows the distributions of the pair opening
angle, $\theta_{+-}$, normalised to $1/E$ for conversions on argon.
At high energies, they peak at the most probable value,
$\hat{\theta}_{+-}$, computed by Olsen in the high-energy
approximation \cite{Olsen:1963zz}, that is indicated by a vertical
line:
\begin{equation}
\hat{\theta}_{+-} = \gfrac{3.2\,m c^2}{E} 
,
\label{eq:Olsen}
\end{equation}
where $m$ is the mass of the pair lepton. Here $m = m_\mu$ and
\begin{equation}
 \hat{\theta}_{+-} \approx \gfrac{338.1\,\mega\electronvolt}{E}
 .
\label{eq:Olsen:mu}
\end{equation}

\section{Conclusion}

I have adapted the sampler of the Bethe-Heitler differential cross
section that I had written for gamma conversions to $e^+e^-$ pairs
\cite{Bernard:2018hwf} to the generation of gamma conversions to
$\mu^+\mu^-$, by simply tuning the parametrisations of the bounds for
variable $x_1$ (Sect. \ref{sec:bounds}) and of the maximum
differential cross section (Sect. \ref{sec:maxdif}).
Verifications of various distributions obtained by the algorithm have
been performed successfully from threshold to PeV energies
(Sect. \ref{sec:dist}).

The Geant4 physics model corresponding to the present algorithm is
about to be made available to users in release 10.6
\cite{Hrivnacova:CHEP2019}.

\section{Acknowledgments}

I'd like to express my gratitude to Vladimir Ivantchenko who brought
the need for a correct low-energy $\gamma \to \mu^+\mu^-$ event
generator to my attention and to Igor Semeniouk and Mihaly Novak for
stimulating discussions.

\end{document}